\documentclass{article}
\usepackage{amsmath,amssymb}
\usepackage[margin=1in]{geometry}
\usepackage{amsthm}
\usepackage{mathtools}
\usepackage{algorithm}
\usepackage[noend]{algpseudocode}
\usepackage{enumerate}
\usepackage{xcolor}
\usepackage{siunitx}
\usepackage{natbib}
\usepackage{caption}
\usepackage{graphicx}
\graphicspath{ {./images/} }

\usepackage{array}      
\usepackage{multirow}
\newcommand\mc[1]{\multicolumn{1}{c}{#1}} 

\newcounter{problem}[section]

\theoremstyle{plain}

\newtheorem*{theorem*}{Theorem}
\newtheorem*{lemma*}{Lemma}

\theoremstyle{definition}

\newtheorem*{question*}{Question}

\theoremstyle{remark}

\newtheorem*{notation*}{Nt}

\newcommand{\eps}{\varepsilon}

\newcommand{\ignore}[1]{}  

\usepackage{hyperref}

\title{Hierarchies of No-regret Algorithms}
\author{Ruqing Xu\thanks{All authors contributed equally to this work.} \and Eric Yachbes\footnotemark[1] \and James Zhang\footnotemark[1]}

\date{December 13, 2024}

\begin{document}

\maketitle

\begin{abstract}
    Our paper studies the setting of players using no-regret algorithms in various two-player games. We address whether having stronger regret guarantees or playing against an opponent with weaker regret guarantees yields higher utilities for the player in question. We consider a hierarchy of algorithms from weakest to strongest: uniform random play, no-regret, and no-swap-regret. We find, counterintuitively, that in many games, no-swap-regret is a worse choice for players (and gives better utility for their opponents). We find the root cause of this phenomenon to be a difference in effective learning rate between the two algorithms, where the no-swap-regret algorithms learn $N$ times slower than no-regret algorithms. To address this, we attempt to equalize learning rates, leading to closer utility between no-regret and no-swap-regret players. Finally, we show that for certain random games with $7$ actions per player, no-swap-regret algorithms can perform noticeably better than no-regret algorithms in a manner that cannot be explained away by unfairly adjusted learning rates.
    
\end{abstract}

\section{Introduction}
Two questions naturally emerge in strategic interactions, including competitive games, wars, and business endeavors. First, a player may consider, if I can choose my opponent, do I benefit from having a weaker opponent? Second, a player may wonder, if I can improve my ability, do I benefit from being a stronger player? 

In zero-sum games like Chess, Go, and war, the answer is intuitive: Yes and Yes. However, the answer ceases to be obvious in situations where coordinated actions can bring better payoff to both players, but those actions require higher-level thinking ability of the players to achieve. For example, economic agents with forward looking abilities can maintain coordination in Repeated Prisoners' Dilemma with reward-and-punishment strategies like grim-trigger. However, agents without the ability to contemplate the influence of today's action on future payoffs are not able to sustain the coordination outcome.

We are interested in asking those questions in the context of algorithmic game theory. How does the payoff to an algorithmic player change when the opponent becomes ``stronger'' or when the player itself becomes ``weaker''? Specifically, we define a player to be ``stronger'' when the player uses an algorithm with stronger regret guarantees. For example, we think that a no-swap-regret algorithm is stronger than a no-regret algorithm, which is in turn stronger than a uniform random algorithm.



Much to our surprise, in all classes of games we experiment, we find that the player's payoff does not change monotonically when we change the opponent's (or the player's) algorithm from uniform random to no-regret, and from no-regret to no-swap-regret. Specifically, changing the opponent from no-regret to no-swap regret  usually increases one's payoff, and changing the player itself from no-regret to no-swap-regret decreases one's payoff. We find this consistent pattern for a wide range of games including different types of auctions, pricing games, and classic matrix games. 

Upon further investigation, we find that the root cause of this surprising phenomenon is the difference in learning rate between the two algorithms. We build no-swap-regret algorithms using the reduction as in \citet{blum2007external}, meaning that the algorithm uses $N$ no-regret algorithms as its subroutines (where $N$ is the number of actions). Specifically, the losses that the no-swap-regret algorithm obtains are divided and distributed to those subordinate algorithms. As a result, the master algorithm learns roughly $N$ times more slowly, at least at the start when the losses are close to equally distributed. 

We have tried a two-pronged approach to address this problem. First, we experiment with amplifying the learning rate parameter of the no-swap-regret algorithm by $N$ times to ``match'' that of its no-regret partner, which gives us some promising results. Second, we attempt to find games where the difference in the algorithms overwhelms the difference in the learning rates. Using a random matrix generation method, we ask the algorithm to flag games in which playing no-swap instead of no-regret against a no-regret opponent gives a better payoff. We fail to do so in smaller sized matrices where $n \in \{2,3,4,5\}$. but we discover $7 \times 7$ matrices where a noticeable difference in utilities is present. 


The paper is structured as follows. Section \ref{sec:lit} connects the paper to the existing literature. Section \ref{sec:methodology} explains our choice of the algorithms, games, and the set-up of our experiments. Section \ref{subsec:unif_learningrate} presents results when we set the same numerical learning rate for the algorithms and the problem with it. Section \ref{subsec:learningrate_adj} presents the case for the learning rate correction. Section \ref{subsec:random_matrix} employs the random matrix generation procedure. Section \ref{sec:discussion} discusses the current results and looks into future directions.

\section{Literature}\label{sec:lit}

This paper connects to the classic literature on algorithms with various regret guarantees.
Specifically, we use the multiplicative weights (MW) algorithm as our no-regret player \citep{arora2012multiplicative}, and the reduction in \citet{blum2007external} to build our no-swap-regret player from MW algorithms. However, there has been limited understanding of how algorithms with different regret guarantees compare and what outcomes arise when they play against each other in general games.

On the other hand, this paper connects to work in ``algorithmic collusion.'' This line of work shows that when learning algorithms (e.g., no-regret algorithms, Q-learning) play against each other, they tend to converge to a collusive outcome rather than a Nash equilibrium \citep{asker2021artificial, banchio2022artificial, arunachaleswaran2024algorithmic}. 
However, this research typically considers scenarios where two or more identical algorithms play against each other, demonstrating that collusion occurs under these conditions. This paper aims to bridge these two lines of work by considering the gameplay dynamics and outcomes between algorithms with different regret guarantees. 

We make use of two important results in section \ref{subsec:random_matrix} when we try to generate matrix games where the no-swap-regret player does better regardless of learning rate. \citet{10.1145/1374376.1374430} show that the repeated play of no-regret learners converges to Coarse Correlated Equilibrium (CCE), and \citet{blum2007external} show that the repeated play of no-swap-regret learners converges to Correlated Equilibrium (CE), which is a subset of CCE.


The learning rate between algorithmic players turn out to be an important driver of our first set of results. The literature on learning rate is sparse, but to our best knowledge, \citet{zrnic2021leads} explicitly explore the effect of learning rate in a strategic classification game and show that by learning faster, the classifier can reverse the order of play and act as a Stackelberg follower. In this light, our paper provides an arguably more general instance where learning rate affect players' payoffs in general class of games.

Finally, this paper broadly connects to the simplicity and robustness literature in economics \citep{pycia2023theory, li2017obviously, carroll2019robustness}. Simplicity asks that, if agents are constrained in their foresight or the ability of making certain inferences, what can be be achieved by mechanisms subject to these simplicity constraints. Robustness asks that, if certain aspects of the environments or the opponents are unknown, what strategies could give better utilities in the worst case. Under this light, this paper examines simplicity when agents are constrained as learning algorithms and robustness when opponents are known to be algorithmic but with unknown regret guarantees.

\section{Methodology}\label{sec:methodology}

    
    \begin{minipage}[c]{0.5\textwidth}
  \centering
  $\setlength{\extrarowheight}{2pt}
  \begin{array}{rr|c c}
      &  \mc{}      & \multicolumn{2}{c}{\textup{Player 2}} \\
      &       & C & D \\ \cline{2-4}
      \multirow{2}{*}{\textup{ Player 1}} & C             &  (0.9, 0.9) & (0, 1) \\ 
       & D            & (1, 0) &  (0.1, 0.1) \\ 
  \end{array}$
  \captionof{table}{Prisoners' Dilemma}
  \label{tab:PD}
\end{minipage}%
\begin{minipage}[c]{0.5\textwidth}
    \centering
    $\setlength{\extrarowheight}{2pt}
  \begin{array}{rr|c c}
      &  \mc{}      & \multicolumn{2}{c}{\textup{Player 2}} \\
      &       & A & B \\ \cline{2-4}
      \multirow{2}{*}{\textup{ Player 1}} & A             & (1, 0.1) & (0, 0) \\ 
       & B            & (0, 0) &  (0.1, 1) \\ 
  \end{array}$
    \captionof{table}{Battle of the Sexes}
    \label{tab:BoS}
  \end{minipage}
  
Our research consists of two phases: simulation and analysis. For our simulations, we implement a codebase in Java from scratch to model the interactions that different players would have in a variety of games. Specifically, we have implemented players capable of playing uniform random, no-regret (multiplicative weights), and no-swap strategies (reduction to multiplicative weights). Each of these algorithms are implemented in such a way that they can be generalized seamlessly to be played in any arbitrary game. We also implement a variety of games, including single-item auctions (first, second, and all pay), pricing games, and general matrix games (including classic games like Prisoners' Dilemma and Battle of the Sexes as in Table \ref{tab:PD} and \ref{tab:BoS}). All of these implement methods from more general game classes using Java inheritance features in order to avoid code redundancy. All games can be played with any number of players (with the exception of general matrix games, which have been limited to two players) playing any arbitrary strategies and having any arbitrary game-dependent data (such as arbitrary item valuations in auction games or arbitrary learning rates for no-regret algorithms). This allows us to run a variety of games with a variety of players without having to write additional code.

Simulations are run in batches and then averaged in order to determine more general properties of games that could not be determined with a single simulation. Specifically, for any game, we run $100$ simulations of games consisting of $1000$ turns, and we take the average utility and strategy picked over all simulations for each turn. We record this data in JSON files, and pipe them to a Jupyter notebook written in Python to plot the mean utilities and strategies picked on graphs. We also store a heat-map of strategies picked to see the approximate distribution of equilibria that games converge to, which is not necessarily determinable with just knowing means.

We have also set up systems which allow us to run large numbers of random simulations (specifically, simulations over many random utility matrices in matrix games) to look for classes of games with certain hierarchies of no-regret strategies. We detail the results of this brute-force search in Section \ref{subsec:random_matrix}. Part of this process involves factoring out the differences in effective learning rate that occur as a result of using different no-swap regret algorithms to be able to determine that no-swap will converge to a different equilibrium than no-regret, not just converging to the same equilibrium at different rates. We have conjectured a means of adjusting the learning rate which has proved effective in minimizing differences in effective learning rate between no-regret and no-swap in small matrix games.


\section{Results}
\subsection{Uniform Learning Rate}\label{subsec:unif_learningrate}

Here are the results of our simulation. As stated in the previous section, we run 100 simulations for each game and a total of 1000 turns for each simulation. For all auctions, the true value of each player is 1, and bids are discretized into increments of 0.05. The payoff matrices for Prisoner's Dilemma and Battle of the Sexes are given in the previous section. In the first set of simulations, we paired uniform random players, multiplicative weights players with $\eps = 0.5$, and no swap regret players with $\eps = 0.5$ against each other. 

The average utilities per turn (averaged across 100 simulations) for selected games are listed below:

\begin{table}[h!]
\centering
\begin{tabular}{c|ccc}
First-Price Auction & Unif Random    & MW(0.5)        & NoSwap(0.5)    \\ \hline
Unif Random         & $(0.16, 0.16)$ & $(0.13, 0.24)$ & $(0.14, 0.22)$ \\
MW(0.5)             & $(0.24, 0.13)$ & $(0.07, 0.07)$ & $(0.11, 0.11)$ \\
NoSwap(0.5)         & $(0.22, 0.14)$ & $(0.11, 0.11)$ & $(0.14, 0.14)$
\end{tabular}

\end{table}
\begin{table}[h!]
\centering
\begin{tabular}{c|ccc}
Second-Price Auction & Unif Random    & MW(0.5)        & NoSwap(0.5)    \\ \hline
Unif Random            & $(0.34, 0.34)$ & $(0.01, 0.50)$ & $(0.07, 0.47)$ \\
MW(0.5)                & $(0.50, 0.01)$ & $(0.02, 0.02)$ & $(0.41, 0.01)$ \\
NoSwap(0.5)            & $(0.47, 0.07)$ & $(0.01, 0.41)$ & $(0.10, 0.10)$
\end{tabular}
\end{table}

\begin{table}[h!]
\centering
\begin{tabular}{c|ccc}
Prisoner's Dilemma & Unif Random    & MW(0.5)        & NoSwap(0.5)    \\ \hline
Unif Random        & $(0.50, 0.50)$ & $(0.06, 0.55)$ & $(0.06, 0.55)$ \\
MW(0.5)            & $(0.55, 0.06)$ & $(0.11, 0.11)$ & $(0.11, 0.11)$ \\
NoSwap(0.5)        & $(0.55, 0.06)$ & $(0.11, 0.11)$ & $(0.11, 0.11)$
\end{tabular}
\end{table}

\begin{table}[h!]
\centering
\begin{tabular}{c|ccc}
Battle of the Sexes & Unif Random    & MW(0.5)        & NoSwap(0.5)    \\ \hline
Unif Random         & $(0.28, 0.27)$ & $(0.05, 0.50)$ & $(0.05, 0.50)$ \\
MW(0.5)             & $(0.50, 0.05)$ & $(0.49, 0.58)$ & $(0.70, 0.36)$ \\
NoSwap(0.5)         & $(0.50, 0.05)$ & $(0.36, 0.70)$ & $(0.48, 0.58)$
\end{tabular}
\end{table}

In the plots below, for auctions, we plot bids vs. turn number where the bid for a specific turn is averaged over 100 simulations. In other words, we average the bid of turn 0 of all 100 simulations and plot it as the first point, the bid of turn 1 of all 100 simulations and plot it as the second point, etc. For matrix games, we plot the proportion of defects (Prisoner's Dilemma) or option $B$ (Battle of the Sexes) vs. turn number, where the proportion is averaged over 100 simulations.

A few notable observations:

\begin{itemize}
    \item In second-price auctions, a uniform random player gets a significantly worse utility against multiplicative weights and no-swap than against another uniform random player. This is because both multiplicative weights and no-swap converge to the optimal strategy of bidding truthfully in a second-price auction.
    \begin{center}
        \includegraphics[width = 0.3 \textwidth]{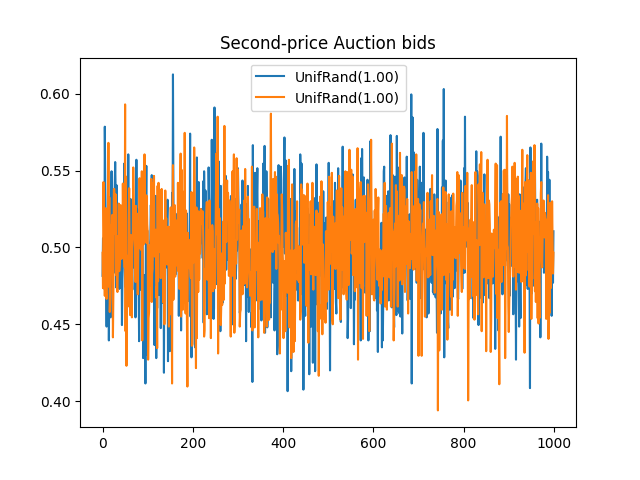}
        \includegraphics[width = 0.3 \textwidth]{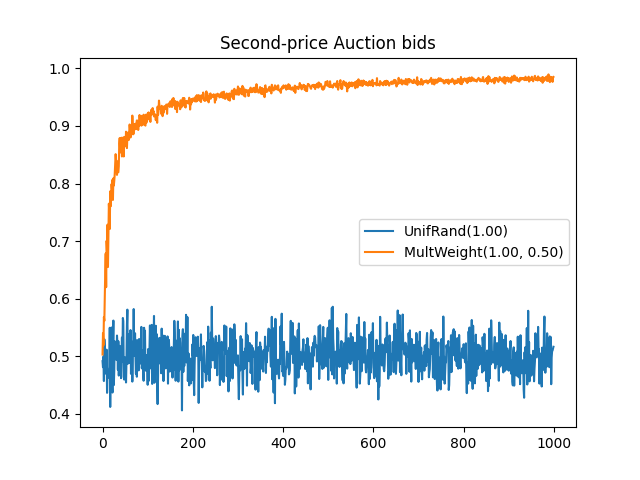}
        \includegraphics[width = 0.3 \textwidth]{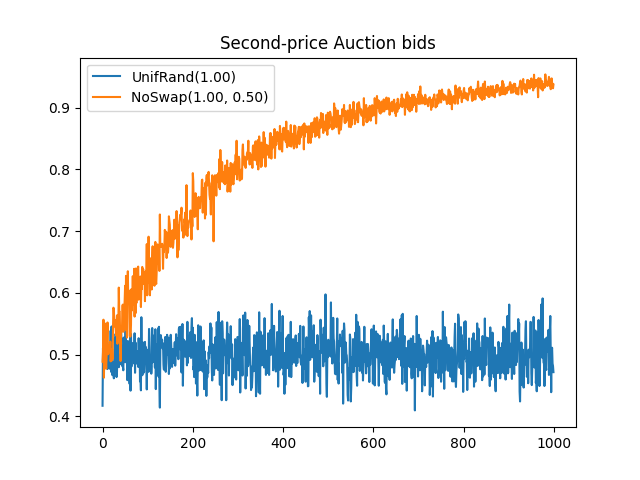}
    \end{center}
    \item For first-price and second-price auctions, a multiplicative weights player gets less utility when playing another multiplicative weights player compared to playing a no-swap player. A no-swap player also gets less payoff when playing a multiplicative weights player compared to playing another no-swap player.

    The plots show that the multiplicative weights player learns the optimal strategy faster than no-swap. This is more obvious in the second-price auction, since after the multiplicative weights player learns the optimal strategy of bidding 1, the other player can't get any positive utility.
    \begin{center}
        \includegraphics[width = 0.3 \textwidth]{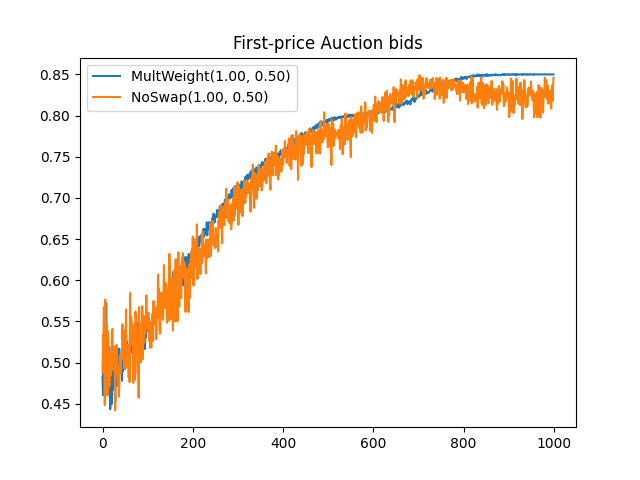}
        \includegraphics[width = 0.3 \textwidth]{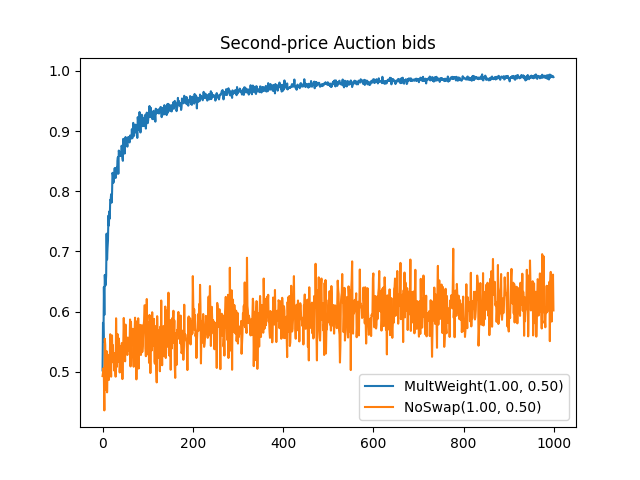}
    \end{center}
    \item For Prisoner's Dilemma, the uniform random player gets a significantly worse average utility against both multiplicative weights and no-swap compared to playing another uniform random player, since both multiplicative weights and no-swap converge to the optimal strategy of defect.
    \item For Prisoner's Dilemma, multiplicative weights and no-swap get similar utilities when paired against each other vs. when paired against another player with the same strategy. The plots show that they both converge to the optimal strategy of defect very quickly, within the first 100 turns.
    \begin{center}
        \includegraphics[width = 0.3 \textwidth]{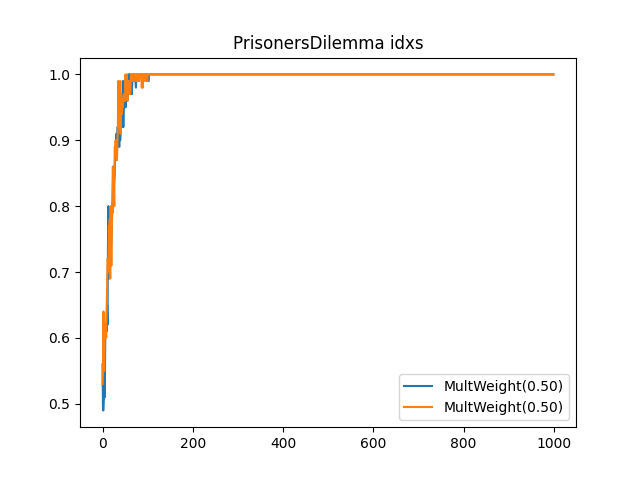}
        \includegraphics[width = 0.3 \textwidth]{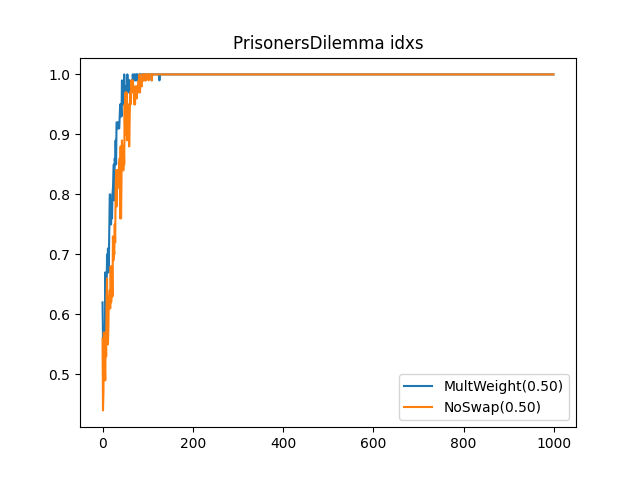}
        \includegraphics[width = 0.3 \textwidth]{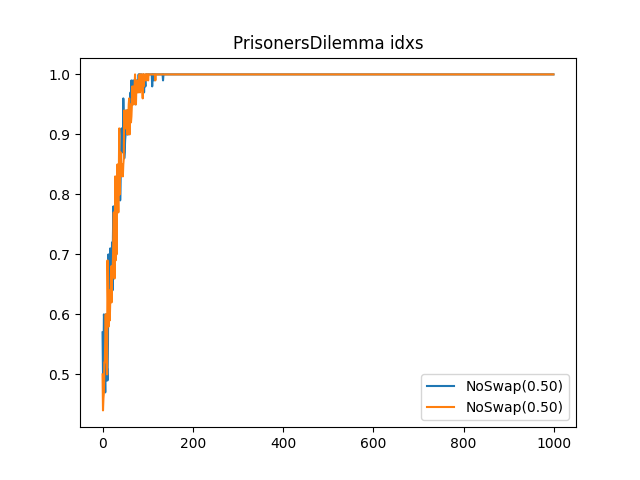}
    \end{center}
    \item For Battle of the Sexes, note that the Nash equilibrium $(A, A)$ is more beneficial to the first player while the Nash equilibrium $(B, B)$ is more beneficial to the second player. Both multiplicative weights and no-swap converged to the Nash that is more beneficial to them when playing against a uniform random player. However, when a multiplicative weights player plays against a no-swap player, it is more likely for the two players to converge to a Nash equilibrium that is more beneficial to multiplicative weights than to no-swap. The plot shows that initially, both players want to drag the other player towards their preferred equilibrium. However, multiplicative weights learns its preferred strategy faster and spends relatively less time playing random strategies. This effectively changes the environment of the no-swap player and makes it a Stackelberg follower. Note that this result seems to be opposite to what the literature commonly suggests, and we discuss this in section \ref{sec:discussion}.
    
    \begin{center}
    \includegraphics[width = 0.3 \textwidth]{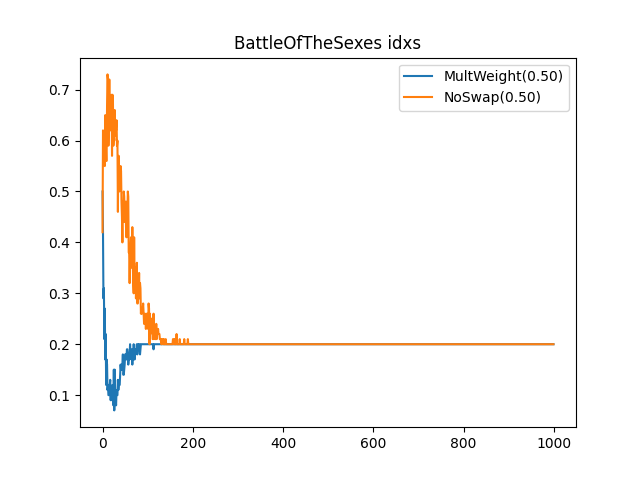}
    \end{center}
\end{itemize}

\subsection{Adjusting the Learning Rate for No-Swap Regret Algorithm}\label{subsec:learningrate_adj}

In light of the results in the previous section, we want to minimize the impact of learning rate in our comparison of algorithms. We propose to adjust the learning rate of the no-swap regret player as follows: If the multiplicative weights player uses learning rate $\eps$ (that is, the weights $w_i$ for each option gets multiplied by $(1 - \eps)^{\ell_i}$, where $\ell_i \in [0, 1]$ is the normalized loss), then the learning rate of the no-swap regret player is $1 - (1 - \eps)^K$ where $K$ is the number of actions (that is, the weights $w_i$ for each option gets multiplied by $(1 - \eps)^{K\cdot \ell_i}$).

For auctions, we set $\eps = 0.1$ for multiplicative weights. We have $K = \frac{1}{0.05} = 20$ and so for no-swap, the learning rate is $1 - (1 - 0.1)^{20} = 0.88$. Here are some results of the simulations after this adjustment. The numbers indicate the average utilities per turn, averaged over 100 simulations:
\begin{table}[h!]
\centering
\begin{tabular}{c|cc}
First-price Auction & MW(0.1)        & NoSwap(0.88)    \\ \hline
MW(0.1)             & $(0.12, 0.12)$ & $(0.09, 0.12)$ \\
NoSwap(0.88)         & $(0.12, 0.09)$ & $(0.08, 0.08)$
\end{tabular}
\end{table}

\begin{table}[h!]
\centering
\begin{tabular}{c|cc}
Second-price Auction & MW(0.1)        & NoSwap(0.88)    \\ \hline
MW(0.1)             & $(0.08, 0.08)$ & $(0.03, 0.16)$ \\
NoSwap(0.88)         & $(0.16, 0.03)$ & $(0.04, 0.04)$
\end{tabular}
\end{table}

For Prisoner's Dilemma and Battle of the Sexes, there are $K = 2$ options for each player, so we set $\eps = 0.5$ for multiplicative weights and $\eps = 0.75$ for no-swap. We run Battle of the Sexes for 1000 simulations rather than 100 since the variations between the simulations are quite large.

\begin{table}[h!]
\centering
\begin{tabular}{c|cc}
Prisoner's Dilemma & MW(0.5)        & NoSwap(0.75)    \\ \hline
MW(0.5)             & $(0.11, 0.11)$ & $(0.11, 0.11)$ \\
NoSwap(0.75)         & $(0.11, 0.11)$ & $(0.11, 0.11)$
\end{tabular}
\end{table}

\begin{table}[h!]
\centering
\begin{tabular}{c|cc}
Battle of the Sexes & MW(0.5)        & NoSwap(0.75)    \\ \hline
MW(0.5)             & $(0.54, 0.54)$ & $(0.53, 0.55)$ \\
NoSwap(0.75)         & $(0.55, 0.53)$ & $(0.56, 0.52)$
\end{tabular}
\end{table}

We can see that for Battle of the Sexes, adjusting the learning rate makes the utilities of multiplicative weights and no-swap much closer to each other. From the plots of which options they chose, it also seems that the two algorithms are learning at roughly the same rate. However, for auctions, multiplicative weights seem to now learn slower than no-swap according to the diagrams, so our adjustment rule works well for $K = 2$ but doesn't work very well for $K = 20$. 

\begin{center}
    \includegraphics[width = 0.3 \textwidth]{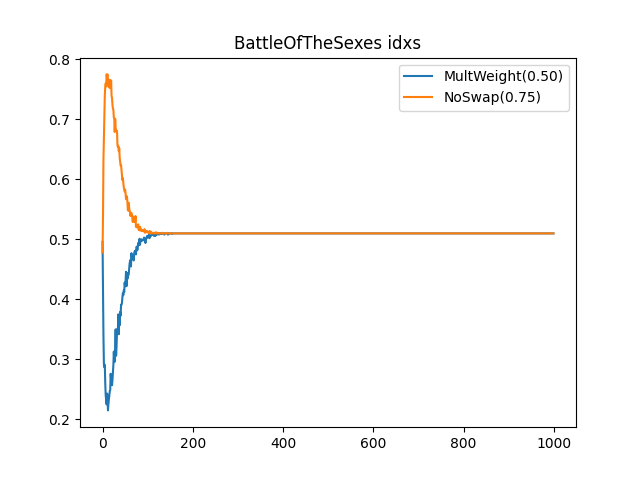}
    \includegraphics[width = 0.3 \textwidth]{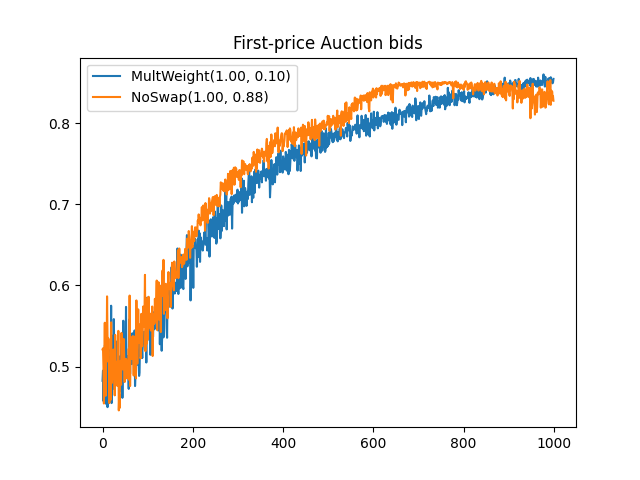}
    \includegraphics[width = 0.3 \textwidth]{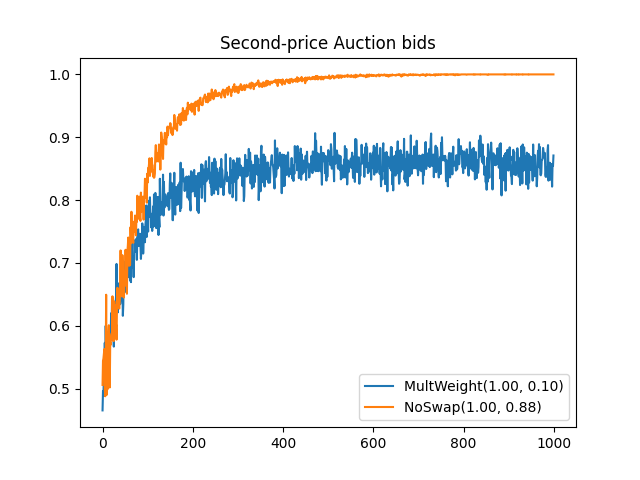}
\end{center}

There is also an abnormal dip in the average bids of no-swap in later turns in first-price auctions. This may be interpreted as the no-swap player trying to exploit the low learning rate of multiplicative weights, so they bid lower than before to pay less and still win the auction.

\subsection{Preliminary Results on Random Matrix Generation}\label{subsec:random_matrix}

Another proposal to circumvent the impact of learning rates is to find games in which it would be better to play no-swap instead of no-regret against a no-regret opponent. To do this, we look specifically at finding matrix games where this is the case. We generate random utility matrices of various sizes with entries picked in a uniform random distribution $[0,1)$. The results are that no significant ($> 0.1$) differences in utility are present in matrices of sizes $n \times n$ where $n \in \{2,3,4,5\}$. However, we discover $7 \times 7$ matrices (although these seem to be sparse among the set of all $7 \times 7$ matrices) where a noticeable difference is present. We test this difference over numerous simulations and determine that such differences are consistently present. For these simulations, the learning rate for no-regret (multiplicative weights) is $\eps = 0.2$ and the learning rate for no-swap is $\eps = 1 - (1 - 0.2)^7 = 0.79$.

Consider the following two payoff matrices, where the payoff of the first player is on the left and the payoff of the second player is on the right. No-regret playing as the second player gets 0.1 more utility per turn than no-swap playing as the second player, when the first player is fixed to be no-regret. 
\[
\begin{bmatrix}
0.365 & 0.003 & 0.066 & 0.183 & 0.336 & 0.969 & 0.888 \\
0.250 & 0.009 & 0.943 & 0.457 & 0.422 & 0.639 & 0.504 \\
0.956 & 0.786 & 0.158 & 0.163 & 0.433 & 0.466 & 0.539 \\
0.848 & 0.543 & 0.841 & 0.056 & 0.116 & 0.241 & 0.336 \\
0.231 & 0.013 & 0.923 & 0.647 & 0.940 & 0.450 & 0.827 \\
0.486 & 0.321 & 0.723 & 0.862 & 0.475 & 0.934 & 0.344 \\
0.400 & 0.560 & 0.830 & 0.648 & 0.017 & 0.478 & 0.048
\end{bmatrix} \, \begin{bmatrix}
0.756 & 0.563 & 0.075 & 0.360 & 0.325 & 0.484 & 0.006 \\
0.966 & 0.197 & 0.756 & 0.192 & 0.243 & 0.491 & 0.391 \\
0.679 & 0.569 & 0.873 & 0.019 & 0.975 & 0.731 & 0.808 \\
0.646 & 0.751 & 0.379 & 0.290 & 0.336 & 0.808 & 0.560 \\
0.192 & 0.239 & 0.558 & 0.375 & 0.674 & 0.044 & 0.877 \\
0.434 & 0.221 & 0.716 & 0.071 & 0.859 & 0.807 & 0.085 \\
0.874 & 0.265 & 0.351 & 0.696 & 0.856 & 0.325 & 0.111
\end{bmatrix}
\]

For the following two payoff matrices, no-swap playing as the second player gets 0.1 more utility per turn than no-regret playing as the second player, when the first player is fixed to be no-regret. We use the same learning rate for these two experiments.

\[
\begin{bmatrix}
    0.869 & 0.498 & 0.005 & 0.766 & 0.339 & 0.630 & 0.962 \\
0.982 & 0.735 & 0.343 & 0.401 & 0.203 & 0.509 & 0.107 \\
0.967 & 0.154 & 0.010 & 0.528 & 0.922 & 0.382 & 0.014 \\
0.989 & 0.994 & 0.378 & 0.860 & 0.011 & 0.044 & 0.914 \\
0.520 & 0.491 & 0.726 & 0.173 & 0.950 & 0.753 & 0.598 \\
0.444 & 0.918 & 0.181 & 0.856 & 0.006 & 0.316 & 0.285 \\
0.994 & 0.280 & 0.621 & 0.248 & 0.186 & 0.538 & 0.811
\end{bmatrix} \, \begin{bmatrix}
    0.332 & 0.130 & 0.228 & 0.515 & 0.065 & 0.425 & 0.445 \\
0.591 & 0.786 & 0.294 & 0.220 & 0.669 & 0.941 & 0.588 \\
0.051 & 0.383 & 0.530 & 0.608 & 0.556 & 0.169 & 0.211 \\
0.177 & 0.205 & 0.542 & 0.890 & 0.812 & 0.846 & 0.690 \\
0.055 & 0.584 & 0.127 & 0.103 & 0.733 & 0.442 & 0.996 \\
0.666 & 0.695 & 0.369 & 0.168 & 0.352 & 0.481 & 0.701 \\
0.801 & 0.182 & 0.855 & 0.364 & 0.699 & 0.621 & 0.563
\end{bmatrix}
\]

It is unclear as of now whether the differences in utilities are a result of unfairly adjusted learning rates or if they are due to no-swap converging to a different equilibrium than no-regret. However, the ability to find matrices where playing no-swap has an advantage and matrices where it doesn't (even when learning rates are fixed to be the same between both games) indicates that there is a difference between these two games that cannot be explained solely based on an unfairly adjusted learning rate.

It may be that even for fixed $\eps$'s and opponents, no-swap regret learns faster on some payoff matrices than on others. We have seen above in Sections \ref{subsec:unif_learningrate} and \ref{subsec:learningrate_adj} that changing $\eps$ or the opponent affects how fast an algorithm converges to an equilibrium and which equilibrium it converges to (as in the Battle of the Sexes game). This example shows that the rate of convergence to an equilibrium may also depend on the payoff matrix. 

\section{Discussion and Future Directions}\label{sec:discussion}
\paragraph{Choice of algorithms and learning rates.} Probably the biggest takeaway from this paper is that when comparing the performance of different algorithms, the inherent learning rate difference can be a driver of results in many generic game settings. Given this, one approach is trying to equalize the learning rates of different algorithms or finding games where the differences in the algorithms overwhelms that of learning rates. An equally valid approach, however, is to recognize the role of learning rate in algorithmic performance and choose algorithms with that in mind. At the very least, this paper once again points to the importance of the understudied topic of learning rates and calls for future exploration.

\paragraph{Learning rates and the Stackelberg equilibrium.} A common belief in the literature states that when an agent updates slower than its opponent, it acts effectively as a Stackleberg leader. However, our result on the Battle of the Sexes game paints a more nuanced picture: the slower no-swap player acts as a follower and gets a worse utility. The reason is that both algorithms start with picking strategies at random, and the no-regret player gets out of the uniform random stage faster and starts to take almost exclusively its preferred action. By this point, the still-learning no-swap player can only adapt to this environment and follows the no-regret player. This would not happen, for example, if both algorithms' default start is heavily taking its preferred action.

\paragraph{Random matrix generation.} We know from existing literature that the play of no-regret algorithms converges to Coarse Correlated Equilibrium (CCE) while that of no-swap-regret algorithms converges to Correlated Equilibrium (CE). We can manually construct games where CEs $\neq$ CCEs, but the CCEs are often unnatural for mean-based algorithms like MW and no-swap to converge to. Results from section \ref{subsec:random_matrix}, although inconclusive, point towards a conjecture: there is practically no difference in a set of CCEs and CEs that the algorithms can converge to in small matrix games. Extensively testing and formalizing this claim can be a future direction of this paper.

\paragraph{Games in more complex environments.} All the games we have considered so far are full-information games. Games with uncertainty about the opponent or the environment, modeled as Bayesian games, can pose more challenges for the learning algorithm to converge to an equilibrium. Recent work \citep{ahunbay2024uniqueness} indeed finds that without some strong assumptions on priors and allowable strategies, no-regret learners may not learn the Bayesian Coarse Correlated Equilibria. Therefore, this could be a promising class of games to look into if the random matrix generation approach does not deliver provable results.








\bibliographystyle{chicago}
\bibliography{bibliography}

\clearpage 

\appendix

\end{document}